# Do I Even Need a Multilevel Model?

# Negligible Effect Significance Testing for Intraclass Correlation Coefficients

Dayeon Lee[1] and Gregory R. Hancock[1]

[1]Department of Human Development and Quantitative Methodology, University of Maryland

## Author Note

Dayeon Lee, https://orcid.org/0000-0003-2806-9993

Gregory R. Hancock, https://orcid.org/0000-0002-6313-006X

We have no conflicts of interest to disclose.

Correspondence concerning this article should be addressed to Dayeon Lee at 1230 Benjamin Building, 3942 Campus Drive, University of Maryland, College Park, MD 20742-1115, United States. Email: daylee450@gmail.com

We thank Nataly Beribisky and Laura Stapleton for their helpful discussions and valuable comments.

**Abstract**

In applied research, the decision to utilize multilevel modeling (MLM) is commonly guided by comparing a point estimate of the unconditional intraclass correlation coefficient (ICC) against some recommended threshold (e.g., 0.05). This naïve approach, however, fails to account for sampling uncertainty and provides no formal inferential justification for a decision. The current work introduces negligible effect significance testing (NEST; or equivalence testing) for the ICC, proposing a framework that performs this test using a pivotal quantity of the ICC based on the $F$ statistic. Furthermore, we offer guidance to help researchers define "negligible" that utilizes the design effect to characterize a tolerable level of variance inflation. R code illustrating the proposed procedure is provided.

## 1. Introduction

In many disciplines within and beyond the social sciences, data are collected that are hierarchical in nature. This data structure violates the statistical assumption of independence required by conventional regression, as observations may share some similarity due to their common higher order units. Ignoring the dependence among individuals could lead to bias in standard errors and in test statistics, resulting in inflated Type I error rates for between-group effects (Dorman, 2008; Raudenbush & Bryk, 2002) and reduced statistical power for within-group effects (Berkhof & Kampen, 2004; Snijder & Bosker, 2012).

To properly consider this hierarchical data structure and obtain correct tests of the parameter estimates, multilevel modeling (MLM; also, hierarchical linear modeling) is the popular choice (Luke, 2004; Raudenbush & Bryk, 2002; Snijders & Bosker, 2012). MLM partitions variance into within- and between-group components (for two-level models), thereby accommodating dependence and yielding valid standard errors. Although data may have been collected deliberately so as to have a hierarchical structure (e.g., via multi-stage sampling) or have such a structure inherently (e.g., students within classrooms), utilizing an MLM might, in fact, be unnecessary given an insufficient degree of observation dependency in those data (i.e., they are *practically* independent). In this case, simple ordinary least squares regression or traditional multiple regression analysis using appropriately weighted data should provide reasonably accurate estimates of the parameters and standard errors (Hayes, 2006; Thomas & Heck, 2001).

To choose between a single-level analysis and a multilevel model, researchers commonly use the unconditional intraclass correlation (ICC, $\rho$) as a diagnostic for evaluating whether the clustering structure induces sufficient within-group dependence to warrant explicit modeling (e.g., Peugh, 2010). The ICC quantifies the proportion of total variance accounted for by between-group differences. Higher values indicate stronger within-group dependence and, consequently, a greater potential for bias in single-level analyses. Although analytic decisions should ultimately reflect substantive research questions, the ICC provides an empirical basis for determining whether clustering is substantively large enough to warrant MLM or negligible enough to justify simpler approaches. Thus, inference regarding ρ can be viewed as formalizing a first-stage screening step to assess whether the dependency among observations within classes or clusters is negligible enough for a single-level analysis or sufficient to warrant proceeding with a multilevel model.

Although no universally accepted guidelines exist for how much dependency is too much, in practice researchers have routinely adopted pre-specified empirical thresholds to justify model selection. For instance, cutoffs ranging from 0.05 (e.g., Dyer, Hanges, & Hall, 2005) to 0.10 (e.g., Lee, 2000) are frequently cited in the educational and psychological literature as sufficient evidence to warrant employing MLM. However, as the assumption of independence is a population-level assumption, the decision really should be based on the status of dependence in the population (ρ), not merely a sample-based description. In this spirit, an approach using a minimum effect (or simple) hypothesis test on the ICC was proposed by Raykov (2011). This framework uses confidence intervals to test whether the ICC is *at least* as large as a prespecified minimum value of substantive interest ($\rho^*$). The hypothesis structure is formulated as $H_0$: $\rho \leq \rho^*$ versus the alternative $H_A$: $\rho > \rho^*$. In this framework, rejecting the null hypothesis provides affirmative statistical evidence that the ICC is meaningfully large, offering a robust justification for employing MLM.

While this method is an improvement over merely inspecting a sample-based ICC estimate, its inferential logic warrants careful interpretation in that a nonsignificant result does not technically provide evidence for a negligible ICC; it merely represents a failure to find evidence of a *substantial* one. Using this strategy, any under-powered such test would also lead to the default conclusion that MLM is *not* warranted; however, as Carl Sagan and many others have long cautioned: “absence of evidence is not evidence of absence” (e.g., Altman & Bland, 1995, p. 485). Indeed, applied researchers should be approaching this from the opposite perspective. As Raykov (2011) himself correctly noted, researchers are typically more interested in determining “whether the ICC itself could be considered small enough to proceed perhaps with single-level analyses” (p. 74). Thus, the critical question is not whether the population ICC is larger than some sufficiently small nonzero threshold in order to justify the complexity of MLM, but rather whether the population ICC is *small* enough to justify the simplicity of a single-level analysis.

To provide a statistically rigorous justification for researchers who seek to proceed with a more parsimonious single-level approach, this paper proposes testing the ICC using the framework of *equivalence testing*, also known as *negligible effect significance testing* (NEST) (e.g., Anderson & Hauck, 1983; Beribisky & Cribbie, 2026; Lakens, Scheel, & Isager, 2018; Rogers et al., 1993; Schuirmann, 1987; Seaman & Serlin, 1998; Tryon, 2001; Wellek, 2010; Westlake, 1976). While the term “equivalence testing” has been used more commonly, the term NEST is conceptually broader and often more appropriate, particularly in the context of parameters such as the ICC. This is because our interest lies in evaluating whether the ICC is negligibly small for practical purposes, rather than whether two quantities (i.e., the population ICC and the equivalence boundary/threshold) are strictly equivalent. Therefore, we will use the term NEST for the remainder of the article.

Inverting the minimum effect hypothesis testing or the traditional null hypothesis significance testing (NHST), NEST is explicitly designed to provide statistical evidence for the *absence* of a meaningful effect (i.e., evidence of a negligible effect). While the choice between MLM and single-level framework should primarily be guided by the research question (e.g., whether cluster-specific inference is required; McNeish, Stapleton, & Silverman, 2017), the NEST approach serves as a critical diagnostic for the validity of statistical inferences. Specifically, it informs researchers whether the clustering effect is sufficiently negligible to proceed with a single-level analysis (with naïve standard errors), or if the dependency is strong enough to induce bias, necessitating a MLM or robust variance estimation.

## 2. Negligible effect significance testing (NEST)

In this section, we briefly review the NEST approach, which will be applied to evaluate ICCs. As noted earlier, when the research objective is to determine whether an effect is sufficiently small so as to be deemed negligible (e.g., a mean difference or correlation), the NHST framework in its traditional implementation is ill-suited as it cannot provide direct evidence to support a null or negligible effect. Alternatively, the NEST framework directly evaluates evidence of a negligible association by inverting the hypothesis structure. Specifically, NEST posits the null hypothesis as stating that the effect is non-negligible, that is, greater than or equal to a minimally meaningful equivalence bound (EB) effect size, $\delta$. The alternative hypothesis, conversely, asserts that the effect is small enough to be considered negligible because it falls within a predefined negligible effect interval (one-sided, true population effect < $\delta$; two-sided, $-\delta$ < true population effect < $\delta$). By rejecting the null hypothesis in favor of the alternative, researchers can directly and statistically conclude that an effect is negligible (Beribisky & Cribbie, 2026; Rogers et al., 1993).

Depending on the nature of the research questions, as telegraphed above, the NEST approach in general can involve either a single one-sided test or two one-sided tests (see Beribisky & Cribbie, 2026). Consider, for example, the scenario of comparing the means of three groups (i.e., ANOVA). Here, a researcher may be interested in whether the mean differences among three groups are overall so small that they may be considered negligible or practically unimportant (thereby warranting a single one-sided test). To answer this question, we can use the effect size ($\eta^2 = SS_{\text{between}}/SS_{\text{total}}$), which represents the proportion of variance in a dependent variable that is explained by the group membership. To use the NEST approach, we first need to set the EB, $\delta$, which serves as a threshold separating negligible from non-negligible effects. Because the effect size $\eta^2$ is bounded by zero, the null and alternative hypotheses for the NEST procedure using $\delta$ (assuming that $\delta$ is set to a specific value) can be written generically as follows:

$$H_0: \eta^2 \geq \delta$$

$$H_1: \eta^2 < \delta.$$

Here, $\delta$ corresponds to the upper bound of a negligible effect interval. To examine this one-sided null hypothesis, a confidence interval (CI) approach developed by Westlake (1981) can be used. To test the above one-sided negligible effect hypothesis, a $100(1-2\alpha)\%$ CI (e.g., 90% CI when $\alpha=0.05$) is used rather than a $100(1-\alpha)\%$ CI because this approach properly preserves the nominal Type I error rate of $\alpha$ for the one-tailed test (i.e., the lower tail of the CI is effectively ignored). This approach follows the same logic as the two one-sided tests (TOSTs, see Schuirmann, 1987) procedure. Thus, using a $100(1-2\alpha)\%$ CI and utilizing only the upper bound will maintain the overall nominal $\alpha$ level (Beribisky & Cribbie, 2026). Specifically, the upper bound of this CI is compared with the equivalence bound $\delta$. If the upper bound of the CI is less than δ, the null hypothesis is rejected, providing statistical evidence that the effect in question, in this case between-group variability, is negligible. Conversely, if the upper bound is greater than or equal to δ, we fail to reject the null hypothesis, and conclude that there is not enough evidence to deem the effect negligible.

## 3. Deriving the upper bound of the confidence interval for the ICC

In the case of the ICC, because it is bounded by zero, the associated NEST approach will also be one-sided, and will require the appropriate sampling distribution in order to conduct the statistical test. Although Raykov (2011) derived an asymptotic approximation to the ICC sampling distribution using the delta method to estimate its standard error, this is technically optimal only under large sample conditions. In much multilevel research within the social and behavioral sciences, however, researchers often face situations where either the number of clusters (level 2 units) or the cluster sizes (level 1 units within clusters) are limited, rendering the asymptotic approximation potentially unreliable (McNeish & Stapleton, 2016).

Fortunately, deriving the complete sampling distribution of a given statistic (here, the ICC), either theoretically or empirically, is not a prerequisite for using the NEST approach. Rather, we only need the boundary values of the CI, thus being operationally equivalent to procedures based on the full sampling distribution while circumventing potentially extensive mathematical derivations (Westlake, 1981). Accordingly, in this section, we derive the *pivotal quantity* (Casella & Berger, 2002) of the sampling distribution of the ICC to obtain the CI's upper bound. A pivotal quantity, a concept originating from Fisher (1930) who used the term *inverse probability*, is a function of observations and unobservable parameters such that the function's probability distribution does not depend on the unknown parameter(s). Once a pivotal quantity is identified, its known distribution can be used to construct a CI by finding appropriate quantiles that serve as the boundaries. For example, consider a normal distribution $N(\mu, \sigma^2)$. When conducting a hypothesis test on the mean, we typically use the sampling distribution of the sample mean, which follows a normal distribution with mean $\mu$ and the variance of $\sigma^2/n$. However, we cannot directly use this distribution, as it depends on the unknown parameter $\mu$. In other words, we cannot build the CI because the distribution can be changed depending on $\mu$. Rather than using the sampling distribution directly, we use a $z$ statistic, defined as $z = \frac{\bar{X}-\mu}{\sigma/\sqrt{n}}$. Here, this entire expression of $z$ is the pivotal quantity, which follows a standard normal distribution $N(0, 1)$. The key property of this pivotal quantity is that its distribution is completely known and does not depend on $\mu$, the parameter about which we wish to make an inference. Because this $z$ follows the well-known standard normal, we can use it to build the confidence interval. Specifically, as we know that

$$P\left(-1.96 \leq \frac{\bar{X}-\mu}{\frac{\sigma}{\sqrt{n}}} \leq 1.96\right) = 0.95,$$

this can be rearranged to obtain the familiar 95% CI:

$$P\left(\bar{X} - 1.96\frac{\sigma}{\sqrt{n}} \leq \mu \leq \bar{X} - 1.96\frac{\sigma}{\sqrt{n}}\right) = 0.95.$$

More formally, if the $Q(Y, \theta)$ is a pivotal quantity with a known distribution where $Q(.)$ is a pivot quantity and $Y$ is data, and $\theta$ is the unknown parameter, we can determine constants $a$ and $b$ such that $P(a < Q(Y, \theta) < b) = 1-\alpha$ (assuming a $100(1-\alpha)\%$ CI). By algebraically inverting this, we can isolate the parameter $\theta$ to obtain confidence interval boundaries. This approach allows us to construct exact CIs without relying on asymptotic approximations, as illustrated below in the context of the ICC.

***Model specification and ICC definition***

Assuming a two-level multilevel model, the unconditional model can be written as:

$$\begin{aligned} &\text{Level 1: } Y_{ij} = \beta_{0j} + r_{ij}, \text{ where } r_{ij} \sim N(0, \sigma^2) \\ &\text{Level 2: } \beta_{0j} = \gamma_{00} + u_{0j}, \text{ where } u_{0j} \sim N(0, \tau_{00}) \end{aligned} \quad (3.1)$$

where $\tau_{00}$ is the error variance among clusters (i.e., between-group variance), and $\sigma^2$ is the error variance among observations (i.e., within-group variance). Here, the ICC ($\rho$) can be computed as:

$$\rho = \frac{\tau_{00}}{\tau_{00} + \sigma^2}. \quad (3.2)$$

This definition of the ICC as the proportion of total variance explained by the between-group variance is conceptually equivalent to the Case 1 (one-way random effects) model specifically designated as ICC(1,1) by Shrout and Fleiss (1979). An ICC of $\rho=0$ represents that there is zero variance attributed to between cluster differences, indicating the independence of the observations (no cluster effect). Conversely, an ICC approaching 1 indicates that the majority of the variance in the data is attributable to differences among clusters. In general, ICC has a value such that $0 \leq \rho \leq 1$, and can inform the methodological decision between applying a conventional single-level model or a more complex MLM.

***Derivation of the pivotal quantity***

The *F*-based pivotal quantity used below follows from well-established results in analysis of variance (ANOVA) and ICC inference (e.g., Arnold, 1981; Box, 1954; Donner, 1986; McGraw & Wong, 1996). We present the derivation not as a new distributional result, but to show how the NEST framework can be applied to the ICC.

To start, the ICC in Eq. (3.1) can be rewritten using the population within-group (error, E) variance ($\sigma_E^2$) and the between-group variance ($\sigma_B^2$):

$$\rho = \frac{\sigma_B^2}{\sigma_B^2 + \sigma_E^2}. \tag{3.3}$$

In the ANOVA framework, we estimate these between- and within-group variances using the mean square between groups ($MS_B$) and the mean square within groups ($MS_W$), respectively. $MS_W$ isolates individual-level residual variance because it is computed from within-group deviation ($Y_{ij} - \bar{Y}_{.j}$), removing all between-group differences. $MS_B$, in contrast, reflects both the true between-group variance ($\sigma_B^2$) and the sampling error related to estimating each group mean ($\frac{\sigma_E^2}{n_j}$). For an unbalanced design, $n_0$ summarizes the unequal group sizes into the average group size adjusted, so that the contribution of between-group variance ($\sigma_B^2$) to the expected $MS_B$ is properly weighted. Averaging the sampling error terms ($\frac{\sigma_E^2}{n_j}$) across groups produces the overall $\sigma_E^2$ component in E($MS_B$), while the between-group variance contributes $n_0\sigma_B^2$. $n_0$ is computed as:

$$n_0 = \frac{1}{J-1}\left(N - \frac{\sum_{j=1}^{J} n_j^2}{N}\right),$$

where $N$ is the total number of subjects (individuals). If it is a balanced design, then $n_0$ would equal $n$ (i.e., the number of observations per group). Therefore, the expected mean squares are computed as:

$$\begin{gathered} \mathrm{E}(MS_W) = \sigma_E^2, \\ \mathrm{E}(MS_B) = \sigma_E^2 + n_0\sigma_B^2. \end{gathered} \tag{3.4}$$

Under the standard normality assumption (i.e., $r_{ij} \sim N(0, \sigma^2)$ and $u_{0j} \sim N(0, \tau_{00})$), the sum of squares from which mean squares are derived (i.e., $SS_B$ and $SS_W$) are related to the $\chi^2$ distribution (Searle, 1971; Thomas & Hultquist, 1978):

$$\frac{SS_{\mathrm{W}}}{\sigma_{\mathrm{E}}^2} = \frac{(N-J)MS_{\mathrm{W}}}{\sigma_{\mathrm{E}}^2} \sim \chi^2(N-J)$$
$$\frac{SS_{\mathrm{B}}}{\sigma_{\mathrm{E}}^2 + n_0\sigma_{\mathrm{B}}^2} = \frac{(J-1)MS_{\mathrm{B}}}{\sigma_{\mathrm{E}}^2 + n_0\sigma_{\mathrm{B}}^2} \sim \chi^2(J-1). \quad (3.5)$$

The $F$ statistic is the ratio of two independent $\chi^2$ variables, each divided by its degrees of freedom. Therefore, the ratio of $MS_B$ to $MS_W$ ($MS_B/MS_W$) has a sampling distribution that is a scaled $F$-distribution (Searle, 1971; Thomas & Hultquist, 1978):

$$F_{\mathrm{obs}} = \frac{MS_{\mathrm{B}}}{MS_{\mathrm{W}}} \sim \frac{E(MS_{\mathrm{B}})}{E(MS_{\mathrm{W}})} \cdot F_{(df_{\mathrm{B}}, df_{\mathrm{W}})}. \quad (3.6)$$

Substituting Eq. (3.4) into Eq. (3.6):

$$\frac{MS_{\mathrm{B}}}{MS_{\mathrm{W}}} \sim \frac{\sigma_{\mathrm{E}}^2 + n_0\sigma_{\mathrm{B}}^2}{\sigma_{\mathrm{E}}^2} \cdot F_{(J-1, N-J)}.$$

This simplifies to:

$$\frac{MS_{\mathrm{B}}}{MS_{\mathrm{W}}} \sim \left(1 + n_0 \frac{\sigma_{\mathrm{B}}^2}{\sigma_{\mathrm{E}}^2}\right) \cdot F_{(J-1, N-J)}. \quad (3.7)$$

This can be connected with the population ICC (ρ) by using Eq. (3.3):

$$\rho(\sigma_{\mathrm{B}}^2 + \sigma_{\mathrm{E}}^2) = \sigma_{\mathrm{B}}^2$$
$$\rho\sigma_{\mathrm{E}}^2 = \sigma_{\mathrm{B}}^2 - \rho\sigma_{\mathrm{B}}^2 = \sigma_{\mathrm{B}}^2(1-\rho)$$
$$\frac{\sigma_{\mathrm{B}}^2}{\sigma_{\mathrm{E}}^2} = \frac{\rho}{(1-\rho)}. \quad (3.8)$$

Using Eq. (3.8), Eq. (3.7) can be rewritten as:

$$\frac{MS_{\mathrm{B}}}{MS_{\mathrm{W}}} \sim \left(1 + n_0 \frac{\rho}{(1-\rho)}\right) \cdot F_{(J-1,N-J)}. \tag{3.9}$$

The relation in Eq. (3.9) indicates that the test statistic calculated from the data ($MS_{\mathrm{B}}/MS_{\mathrm{W}}$) follows a scaled version of the standard *F*-distribution, multiplied by a scaling factor $\left(1 + n_0 \frac{\rho}{(1-\rho)}\right)$.

### *Deriving the upper bound of the confidence interval for ρ*

In order to derive the upper bound of a CI for the ICC (ρ), we now need to derive a pivotal quantity. Although the pivotal quantity is a function that contains ρ (such as $Q(Y, \rho)$), its probability distribution does not depend on the value of ρ. In other words, regardless of what the true value of ρ is, the pivot quantity ($Q$) follows the same known distribution (e.g., *F*-distribution with fixed degrees of freedom, in this ICC case). In Eq. (3.9), we need to divide the scaling factor part $\left(1 + n_0 \frac{\rho}{(1-\rho)}\right)$ from both sides:

$$\frac{\frac{MS_{\mathrm{B}}}{MS_{\mathrm{W}}}}{\left(1 + n_0 \frac{\rho}{(1-\rho)}\right)} \sim F_{(J-1,N-J)}. \tag{3.10}$$

In Eq. (3.10), the left side term is the pivotal quantity ($Q$), which is constructed from observable sample statistics ($MS_{\mathrm{B}}$ and $MS_{\mathrm{W}}$) and the population parameter of interest (ρ). Importantly, this pivotal quantity has a distribution that does not depend on any unknown parameters and follows a known *F*-distribution with degrees of freedom of $J-1$ and $N-J$. This property enables the direct derivation of exact CIs for the ICC without requiring knowledge of the true parameter value or reliance on large-sample approximations.

**4. NEST for the ICC**

As previously explained, the NEST approach inverts the hypotheses' structure from traditional NHST, setting the alternative hypothesis to be that of a negligible effect. Thus, rather than seeking evidence that the ICC is large enough to be considered meaningful (thus invoking MLM), the NEST strategy provides a direct statistical tool to answer the practical question that motivates ICC estimation in the first place: "Is the clustering effect small enough to be justifiably ignored?" To this end, the hypotheses can be stated as:

$$\begin{aligned} &H_0: \rho \geq EB \\ &H_1: \rho < EB, \end{aligned} \tag{4.1}$$

where ρ is the ICC and EB is equivalence bound, which represents the threshold for a negligibly small effect. Within the equivalence testing literature, this threshold is referred to as the smallest effect size of interest (SESOI; Lakens et al., 2018) or equivalence margin (Meyners, 2012), and conceptually corresponds to a minimally meaningful effect size (MMES, δ). Specifically, EB in this context serves as the threshold for a minimally meaningful amount of non-ignorable clustering effect. It also represents the maximum tolerable amount of variance inflation that would result from disregarding the hierarchical data structure (a concept we will explain later in this section using design effect). Given that the ICC is bounded by zero, here MMES is set as a single value (EB), making the NEST approach a one-tailed test.

To conduct the NEST approach for the ICC, the null hypothesis is rejected in favor of ICC negligibility when the upper bound of the ICC's CI falls below EB; otherwise, the null hypothesis of a sufficiently large ICC is retained. To reiterate, failing to reject $H_0$ in this context indicates insufficient evidence to conclude that the ICC is negligibly small. In other words, the data do not provide adequate support for the claim that between-group variance is trivial, and thus researchers cannot confidently rule out the need for MLM to account for the hierarchical structure in the data.

### ***Upper bound of the confidence interval for ρ***

Considering that the NEST approach for the ICC is one-sided and seeks evidence that ρ is sufficiently negligible, it is necessary to determine the upper bound of the CI and compare it to the corresponding EB. To this end, returning to Eq. (3.10), let the ratio ($MS_B/MS_W$) be denoted as $F_{obs}$. Then the upper bound of the CI for ρ, denoted $\rho_U$, is found by solving for the value of ρ that corresponds to the smallest plausible value of the pivotal quantity, $Q$. This inverse relation arises because, as shown in Eq. (4.2) below, ρ appears in the denominators, meaning that as ρ increases toward its upper bound ($\rho_U$), $Q$ decreases toward its lower limits:

$$Q = \frac{F_{\text{obs}}}{1 + n_0 \frac{\rho}{(1-\rho)}}. \tag{4.2}$$

Consequently, to find the upper bound $\rho_U$, we must use the smallest plausible value of $Q$, which is defined by the lower critical value of the $F$-distribution, $F_L$, which is the 100(α)-th percentile of the $F$-distribution with $J$−1 and $N$−$J$ degrees of freedom. Here, $F_L$ is defined as:

$$F_{\text{L}} = F_{\alpha, J-1, N-J}.$$

As previously noted, to conduct this one-tailed test at a nominal α level, we adopt the logic from the TOST procedure, which requires using one bound from a 100(1−2α)% CI (e.g., a 90% CI when α=0.05). To find the upper bound of this interval, which is $\rho_U$, we must use the lower α critical value of the $F$-distribution, $F_L$. Eq. (4.3) expresses this relation:

$$F_{\mathrm{L}} = \frac{F_{\mathrm{obs}}}{1 + n_0\left(\frac{\rho_{\mathrm{U}}}{1-\rho_{\mathrm{U}}}\right)}. \tag{4.3}$$

By solving Eq. (4.3) in terms of $\rho_U$, we obtain the upper bound of the 100(1−2α)% confidence interval:

$$\rho_{\mathrm{U}(1-2\alpha)} = \frac{F_{\mathrm{obs}} - F_{\mathrm{L}}}{F_{\mathrm{obs}} + (n_0 - 1)F_{\mathrm{L}}} = \frac{F_{\mathrm{obs}} - F_{\alpha,J-1,N-J}}{F_{\mathrm{obs}} + (n_0 - 1)F_{\alpha,J-1,N-J}}. \tag{4.4}$$

The Eq. (4.4) provides the upper bound of the 100(1−2α)% CI of the ICC. This upper bound is the test statistic for the NEST procedure: if $\rho_U$ falls below the researcher-specified EB, the null hypothesis of non-negligible clustering is rejected, providing formal statistical justification for regarding the hierarchical data structure as benign. Conversely, if the upper bound meets or exceeds EB, the null hypothesis is retained and the default MLM approach remains recommended. This framework thus offers researchers an evidence-based approach to making the methodological decision between single-level and multilevel analyses based on the magnitude of the ICC. It is worth reiterating that while the population parameter for the between-group variance ($\sigma^2_{\mathrm{B}}$) is a theoretically non-negative value (i.e., $\rho \geq 0$), the ANOVA-based estimator can yield negative estimates due to sampling error (i.e., when $MS_W > MS_B$). The pivotal quantity approach used here (Eq. 4.4) relies on the exact distribution of the *F*-statistic and does not truncate the confidence bounds at zero. In the context of NEST, obtaining a negative upper bound ($\rho_U < 0$) creates no logical contradiction. Rather, it provides evidence that the ICC is well below the positive EB, supporting the conclusion of negligible clustering effect.

### *Choosing the EB*

A critical consideration in implementing the NEST approach is the determination of the EB. Although researchers often rely on conventional thresholds for the ICC such as 0.05 in psychology (e.g., Dyer et al., 2005) or 0.10 in education (e.g., Lee, 2000), a more theoretically grounded approach within the MLM framework involves using the design effect (DEFF). The design effect (Kish, 1965) is defined as the ratio of the variance of estimation under a complex (e.g., clustered) sampling design to the variance of estimation under simple random sampling. In other words, it quantifies the degree to which (squared) standard errors are inflated due to the clustered data structure, relative to an independent sampling design. The MLM explicitly adjusts for this inflation by accounting for the hierarchical nature of the data, thereby yielding statistically valid standard errors.

The DEFF of a simple two-stage sample (assuming no stratification) is computed as:

$$\mathrm{DEFF} = 1 + (n_0 - 1)\rho, \tag{4.5}$$

where $n_0$ is an average group size adjusted for the unequal group sizes. A DEFF of 1 indicates completely independent observations, suggesting that MLM is unnecessary. Conversely, when DEFF exceeds 1, it indicates that disregarding the clustered data structure would result in variance inflation (and corresponding underestimation of standard errors). This implies that dependence among observations is non-negligible, potentially warranting the application of a MLM framework.

Because the DEFF is determined jointly by ρ and $n_0$, researchers should also consider the DEFF, rather than the ICC alone, when deciding whether MLM is needed. Even a small ICC can yield a large DEFF if the average cluster size is large, resulting in substantial inflation of standard errors when single-level analyses are applied. Thus, the DEFF provides a more direct and practical rationale for adopting MLM, as it explicitly captures the combined effects of clustering magnitude and cluster size.

Rearranging the Eq. (4.5) for ρ gives:

$$\rho = \frac{\text{DEFF} - 1}{n_0 - 1}. \tag{4.6}$$

In this equation, ρ can be interpreted as the EB, given an acceptable level of DEFF (i.e., a tolerable level of SE inflation), specified by the researcher. By defining how much inflation in variances (or standard errors) can be tolerated, which is reflected in DEFF, researchers can dynamically determine an EB for the ICC that is tailored to their data. For example, if the average cluster size is 22 (assuming equal group size) and the acceptable design effect is set to 1.1 (i.e., 10% inflation in variance of estimation, which corresponds to a 4.88% inflation in the standard error, that is, sqrt(1.1)≈1.0488), the corresponding EB can be calculated as:

$$\text{EB} = \frac{1.1 - 1}{22 - 1} = 0.005.$$

In this case, demonstrating through a NEST approach that the ICC is smaller than 0.005 provides a direct empirical rationale for using a single-level analysis.[1]

This approach is particularly useful because a small ICC does not necessarily imply that MLM is unwarranted (Huang, 2016, 2018). By linking the EB directly to the DEFF, this framework provides applied researchers with a concrete decision criterion. Specifically, the NEST procedure in this context effectively compares the variance inflation incurred by disregarding clustering against the maximum variance inflation the researcher is willing to accept. Although the question of what constitutes an acceptable DEFF (i.e., how much inflation in variance is tolerable) remains open, basing the EB on DEFF offers a theoretically grounded and quantitatively more transparent criterion for defining the threshold used in NEST. In contrast to the conventional ICC-based rule of thumb (e.g., ICC<0.05; Dyer et al., 2005), which may be well-established in certain fields, the DEFF-based criterion offers an alternative approach that allows researchers to set a context-specific, data-adaptive boundary reflecting the actual structure of their data. Researchers are thus free to choose the approach most appropriate for their discipline and research context.

[1] It should also be noted that an acceptable level of DEFF could be defined differently depending on the research questions. For instance, researchers may choose it more conservatively when precise estimation is critical, or more leniently in exploratory analyses (for more discussion, see the Discussion and conclusions section).

### *Equivalence testing and power*

It is important to note that in the NEST approach, the sample size required to achieve a target level of statistical power tends to be larger than in traditional NHST (Lakens, 2017; Rusticus & Lavoto, 2014). This is because the goal of NEST is to reject effects that are larger than a prespecified EB, not to fail to reject the null hypothesis of no effect. This requires reasonably precise estimation so that the CI lies entirely within the negligible effect region. In other words, in the NEST approach for the ICC the CI must be narrow enough for the upper bound to remain below EB, and it depends on both the number of clusters $J$ and the cluster size $n$.

To provide additional practical guidance, we examined the relation between sample size and power. Under a balanced design with $J$ clusters of size $n$, power ($\pi$) can be defined as follows (for readers who are interested in derivation of this, see Appendix A):

$$\pi_{\text{NEST}} = \Pr\left(F_{J-1,N-J} < \frac{1-\rho}{(n-1)\rho+1} \cdot \frac{(n-1)\text{EB}+1}{1-\text{EB}} \cdot F_{\alpha,J-1,N-J}\right).$$

Here, if $\rho<\text{EB}$ (i.e., $H_0$ is rejected), then the first two terms on the righthand side exceed 1, thereby increasing power. If $\rho=\text{EB}$, then power corresponds to $\alpha$ level. Although power is a function of $n$, $J$, EB, and $\alpha$ (assuming $N=Jn$ for the balanced clustering), we only manipulated $n$, $J$, and $\rho$ (i.e., fixed $\alpha$ and EB) in the simulation. The sample size per group ($n$) and the number of clusters ($J$) ranged from 5 to 195 in increments of 10, respectively. The true ICC ($\rho$) is ranging from 0, 0.01, 0,04, and 0.05 to make them symmetric. EB and $\alpha$ are fixed to 0.05, respectively. For the comparison, NHST were conducted with the $H_0$: $\rho=0$ and $H_1$: $\rho>0$. Because EB is 0 in the NHST, power ($\pi$) in NHST is computed as:

$$\pi_{\text{NHST}} = 1 - \Pr\left(F_{J-1,N-J} < \frac{1-\rho}{(n-1)\rho+1} \cdot F_{\alpha,J-1,N-J}\right).$$

The simulation results are presented in Figures 1 and 2. Excluding the case of $\rho=0$ for NHST and the case of $\rho=0.05$ for NEST, for which the power was equal to the $\alpha$ level, power generally increased as both $n$ and $J$ increased, and NHST tended to reach a given target power (e.g., 0.80) somewhat more quickly. For NEST, when $\rho$ was slightly below the EB (e.g., 0.04), increases in $n$ appeared to have a more direct impact on power than increases in $J$. When $\rho$ was far below the EB (e.g., 0 or 0.01), a target power level of 0.80 could be attained even with a relatively small number of clusters if the $n$ was large. In contrast, when the $n$ was very small, a much larger $J$ was required to achieve the same target power, and in some cases (e.g., $n$=5) the target was not reached at all.

When comparing NHST with $\rho=0.01$ to NEST with $\rho=0.04$, NHST generally reached the target power level as $n$ and $J$ increased, except under extremely small cluster sizes. In contrast, for NEST, even substantial increases in both $n$ and $J$ were insufficient to achieve power of 0.80 in any condition considered. These results suggest that when the true effect is moderately separated from the EB, NEST requires slightly larger sample sizes than NHST to achieve given level of power, as noted previously by Lakens (2017). Importantly, when the $\rho$ is very close to the EB, achieving some adequate level of power may require substantially larger sample sizes.

**5. Sample application**

In this section, we demonstrate application of the NEST approach for examining the unconditional ICC to determine whether to employ MLM. For this illustration, we use data from the Early Childhood Longitudinal Studies: 2011 Kindergarten Class of 2011 (ECLS-K:2011; National Center of Education Statistics, 2012), which is publicly available. These data were collected periodically from children's kindergarten year in 2010-2011 until the spring of 2016, by which time most participants had completed the fifth grade. For the illustration, we pose a research question concerning reading attitudes among third-grade students, focusing on one outcome variable measuring reading attitude ("I like reading") using data from Wave 7 (i.e., Spring 2014). Although analyzing the large-scale data such as ECLS-K:2011 requires sampling weights and accounting for stratification and clustering in the sampling design to generalize findings to the target population, we ignored them here for simplicity of illustration. Likewise, although responses to this item were on a 4-point response scale ranging from 1 (not at all true) to 4 (very true), the data were treated as interval scaled for the purposes of this example.

At the selected time point, the sample comprised 12,823 students nested within 2,471 schools, with 5.19 students as the average cluster (school) size corrected for variance (i.e., $N$=12,823, $J$=2,471, and $n_0$=5.19). All analyses are conducted using RStudio software (R Core Team, 2023). The ECLS-K data were imported using the `EdSurvey` package in R. The R code for this illustration is presented in Appendix C.

To start, we can estimate the ICC using the null model, computed below as 0.034:

$$Y_{ij} = \beta_{0j} + r_{ij}, \qquad \text{where } r_{ij} \sim N(0, 0.890)$$

$$\beta_{0j} = \gamma_{00} + u_{0j}, \qquad \text{where } u_{0j} \sim N(0, 0.031)$$

$$\text{ICC} = \frac{0.031}{0.890 + 0.031} = 0.034\,.$$

Next, to use the NEST approach, the EB should be determined. Here, although it can be set at the researcher's discretion, using the DEFF to set EB provides a theoretically grounded rationale. Assume the researcher believes that 10% inflation in the variance can be tolerated; that is, the DEFF would be 1.1. Then, using the Eq. (4.6), the EB can be set as 0.0239, as per the following:

$$\text{EB} = \frac{1.1 - 1}{5.19 - 1} = 0.0239.$$

Then, using this EB, the null and alternative hypotheses can be stated as follows:

$$\text{H}_0: \rho \geq 0.0239$$

$$\text{H}_1: \rho < 0.0239.$$

Next, the upper bound of the CI needs to be computed, and this will be compared with the EB of 0.0239. To this end, the observed $F$ statistic is computed as $F_{\text{obs}} = MS_{\text{B}}/MS_{\text{W}} = 1.054/0.889 = 1.186$. With the nominal level of $\alpha$=0.05, the critical value for the lower 5% tail of the $F$-distribution with 2,470 and 12,822 degrees of freedom is 0.9488. Using all of these values, the upper bound of the CI for the ICC can be computed using Eq. (4.4):

$$\rho_{\text{U}(1-2\alpha)} = \frac{1.186 - 0.9488}{1.186 + (5.19 - 1)0.9488} = 0.046.$$

Because the upper bound of 0.046 is greater than the EB of 0.0239, the null hypothesis of a negligible ICC cannot be rejected at the $\alpha$ level of 0.05. This result indicates that the evidence is insufficient to support conducting a single-level analysis, and instead MLM would be recommended.

As seen, the point estimate of the ICC for this illustration was calculated to be 0.034. Had a researcher adhered to the conventional empirical cutoff of 0.10 by simply comparing this point estimate, they would have reached the possibly misleading conclusion that MLM was unnecessary. Failing to model this clustering effect would result in model misspecification, potentially yielding incorrect statistical inference. However, by implementing the NEST approach for the ICC as proposed in this study, a more robust conclusion may be achieved.

## 6. Discussion and conclusions

In practical research scenarios, determining whether to employ MLM often hinges on an evaluation of the ICC. This evaluation, however, frequently relies on the empirical and conventional thresholds (e.g., 0.05) that are applied without sufficient context, or on researcher's subjective interpretation, as there is no universally accepted cutoff. To introduce more statistical rigor, Raykov (2011) proposed using a minimal effect hypothesis test. This approach was designed to find positive evidence that the ICC is large enough to be considered meaningful, thereby justifying the use of MLM. While this was indeed an important inferential improvement, it set the single-level approach as its default stance. As such, a nonsignificant result from a minimum effect test only shows a failure to find evidence of a large effect; it does not establish that the effect is negligible. This leaves the researcher wondering about being able to use a simpler model rather than MLM in an ambiguous position. In addition, although perhaps less critical, the delta method approach utilized is based on asymptotic theory, which assumes a large sample (i.e., large number of groups and large number of subjects within them), which may or may not be realistic in many social science contexts.

To address these limitations, this study formalizes the application of the NEST framework to assess the ICC. This approach inverts the inferential logic of traditional significance testing by repositioning the hypothesis of a negligible effect from the null to the alternative. Thus, rather than seeking evidence that the ICC is large enough to be considered meaningful, the NEST approach provides direct statistical evidence to help establish that the ICC is sufficiently negligible so as to eschew MLM in favor of a simpler traditional linear model. In addition, this approach utilizes a pivotal quantity to obtain the upper bound of the confidence interval for the ICC, which does not rely on large-sample asymptotic theory and thus makes it more practically applicable.

This study offers two critical innovations for applied researchers. First, it provides a clear and practical pathway for justifying the use of a single-level analysis. Second, and perhaps more importantly, it offers a principled option for the most challenging aspect of any NEST procedure – the justification of the EB. While the choice of minimally meaningful effect size is often criticized as subjective, we propose leveraging the DEFF to be able to set an adaptive EB. The primary statistical consequence of ignoring a non-zero ICC is the inflation of Type I error rates for cluster-level predictors. Here, the DEFF quantifies this inflation of variance of estimation, and in turn, the relevant standard errors. Using this, a researcher can make an a priori decision about the maximum level of variance inflation, and thus the standard error inflation, they are willing to tolerate. While choosing the maximum level of variance inflation is still subjective, it is at least more tied to the actual consequence of clustering. The acceptable level of this inflation depends entirely on the research context. For example, if it is high-stakes research such as clinical trial for new drug, the limit might be set very conservatively (e.g., 3%). On the other hand, if it is more low-stakes research such as exploratory pilot study, one's tolerance might be more liberal (e.g., 15%), as the goal is more about relation exploration rather than precise hypothesis testing. Once this tolerance is set, then the EB for the ICC can be calculated using DEFF, adaptively accommodating the cluster sizes of the data that the researcher has. This approach allows researchers to make quantitative decisions based on the statistical consequences the researcher is willing to accept.

It is worth explicitly reminding ourselves that the exact pivotal quantity in our proposed NEST procedure relies on the standard assumptions of the one-way random effects model, including normality and homoscedasticity. When these assumptions are violated, the $\chi^2$ distributional results for $SS_{\mathrm{W}}/\sigma_{\mathrm{E}}^2$ and $SS_{\mathrm{B}}/(\sigma_{\mathrm{E}}^2 + n_0\sigma_{\mathrm{B}}^2)$ in Eq. (3.5) no longer hold exactly. This may compromise the $F$-distribution approximation and the quality of the pivotal quantity, and in turn, the CI coverage as the CI upper bound may be biased. Regarding this, two clarifications are in order. First, the context in which the proposed NEST is intended for use is one in which the researcher is already entertaining MLM, and thus presumably considers the data to adequately satisfy its assumptions, including normality and homoscedasticity. The proposed NEST for the ICC is an additional diagnostic tool to evaluate whether the multilevel structure of the data is enough to warrant that analysis for which the distributional assumptions are already believed sufficiently tenable. Second, and more fundamentally, the term *exact* in "exact CI" does not imply that this approach is free of distributional assumptions. Regarding the normality assumption, although the ANOVA $F$-statistic is known to be relatively robust to moderate violations of the normality assumption when comparing means (e.g., Glass, Peckham, & Sanders, 1972), that robustness does not extend to interval estimation of the ICC, which is a variance ratio. Because the sampling behavior of variance component estimator depends on the shape of underlying distributions, the resulting distortion does not vanish as the number of clusters grows (Box, 1953; Burch, 2011a, 2011b). Asymptotic interval approaches (e.g., Raykov, 2011) inherit this vulnerability as well, given that their standard errors for variance components are no longer valid under non-normal cluster effects (Verbeke & Lesaffre, 1997). Therefore, the exactness of the proposed interval should be interpreted as exactness under the assumed model at any number of clusters, not as freedom from distributional assumptions.

Further, when the homoscedasticity assumption is violated, it requires one additional caution. If the within-group variance differs across groups, a single population ICC may no longer be the proper target (Goldstein, Browne, & Rasbash, 2002), as the ICC in Eq. (3.3) assumes one common within-group variance (Kasim & Raudenbush, 1998; Leckie et al., 2014). Thus, the problem is not only that the CI may be accurate; the estimand itself may need to be reconsidered. Some methods allow inference on between-group variance components when error variances are unequal (Hartung & Knapp, 2005). In practice, when normality or homoscedasticity is questionable, we recommend reporting the proposed interval as well. For non-normality, possible choices include the transformed bootstrap-*t* interval (Ukoumunne et al., 2003) and the kurtosis-adjusted restricted maximum likelihood (REML) interval (Burch, 2011a; 2011b). When the intervals lead to similar conclusions, the result is more reassuring; when they disagree, researchers should rely more on the robust alternatives.

In addition, it should be noted that when cluster sizes are unequal, summarizing them in the single value $n_0$ makes the derived pivotal quantity an approximate pivot rather than the exact pivot (Donner, 1986; Searle, 1971; Thomas & Hultquist, 1978). Specifically, in cases of extremely unbalanced clustering (e.g., a few very large clusters alongside many small ones), the quality of the approximation deteriorates (Demetrashvili, Wit, & van den Heuvel, 2014; Swiger et al., 1964). To address this, several studies have proposed different approaches to obtain the CI for the ICC under unbalanced designs. Wald (1940) was the first to construct an exact pivotal quantity for the unbalanced one-way random effects models, introducing observation-weighted statistics where the weights depend on the unknown variance ratio itself. While this yields exact

CI bounds, these bounds are defined as roots of nonlinear equations that cannot be solved in closed form, making the method difficult to apply in practice. Harville and Fenech (1985) generalized Wald's approach via an eigenvalue decomposition, deriving an exact pivot but requiring heavy computation such as matrix diagonalization and iterative numerical procedures. They also suggested approximate CIs based on arithmetic or harmonic means, which is essentially the same as using $n_0$ in our approach. Similarly, Thomas and Hultquist (1978) proposed an easily-computed statistic based on the harmonic mean of the group sizes and variance of the group means, whose distribution is well approximated by a $\chi^2$ distribution. More recently, Demetrashvili et al. (2014) proposed two approaches – one based on Satterthwaite's approximation using an *F*-distribution and another based on moment-matching with a Beta distribution – to obtain the approximate CI when the clustering is unbalanced. Therefore, if the clustering is extremely unbalanced, such alternative approaches might be used to obtain the upper bound of the CI for the ICC and subsequently to conduct the NEST.

Finally, as previously mentioned, the proposed approach targets the unconditional ICC as a first-stage screening step to determine whether MLM is warranted. That is, we do not claim that it is directly valid for every covariate-adjusted multilevel model. Nevertheless, the framework can be extended to examine residual dependence in a model with covariates through the conditional ICC (Raudenbush & Bryk, 2002). Specifically, because covariates may explain part of the between- and within-group variation, researchers may wish to ask, "After accounting for the covariates, is residual dependence negligible?" This question can be answered differently depending on the level at which the covariates operate. Level 2 covariates can only reduce the ICC, so a negligibility conclusion carries over; level 1 covariates, on the other hand, can increase the ICC, and using a slightly stricter margin (i.e., an adjusted EB) would preserve the conclusion. Readers interested in further details on this issue are referred to Appendix B.

Looking ahead, while the proposed NEST approach offers potential advantages, room for future methodological work remains. First, regarding the violation of assumptions, more systematic evaluations for the performance of the NEST for the ICC (especially CI construction) would be useful for understanding conditions of severely non-normal and/or heteroscedastic data. Second, the current framework is directly applicable to two-level models. Extending this approach to three-level models is not trivial, as the definition of the ICC becomes more complex (e.g., the proportion of variance at the third level). Future research could explore deriving the pivotal quantity for more complex hierarchical structures. Third, regarding sample size, although we briefly examined power under balanced clustering, Rusticus and Lovato (2014) showed that holding total sample size constant, power for equivalence testing decreases as group sizes become more unbalanced. Therefore, further investigation into the impact of unbalanced clustering on power may be of interest.

Ultimately, this NEST approach for the ICC contributes to more rigorous and transparent statistical inference in multilevel research. By shifting the burden of proof to the researcher to establish a negligible effect, it helps researchers to move beyond arbitrary cutoffs and engage more deeply with the question of practical significance. Our hope is that this work provides applied researchers with a clear and defensible standard for making one of the most fundamental decisions in analyzing multilevel data, encouraging a priori thinking about what magnitude of an effect is *truly* meaningful for their research.

**Figure 1.**

*Power curves for NEST across number of clusters, by true ICC and cluster size*

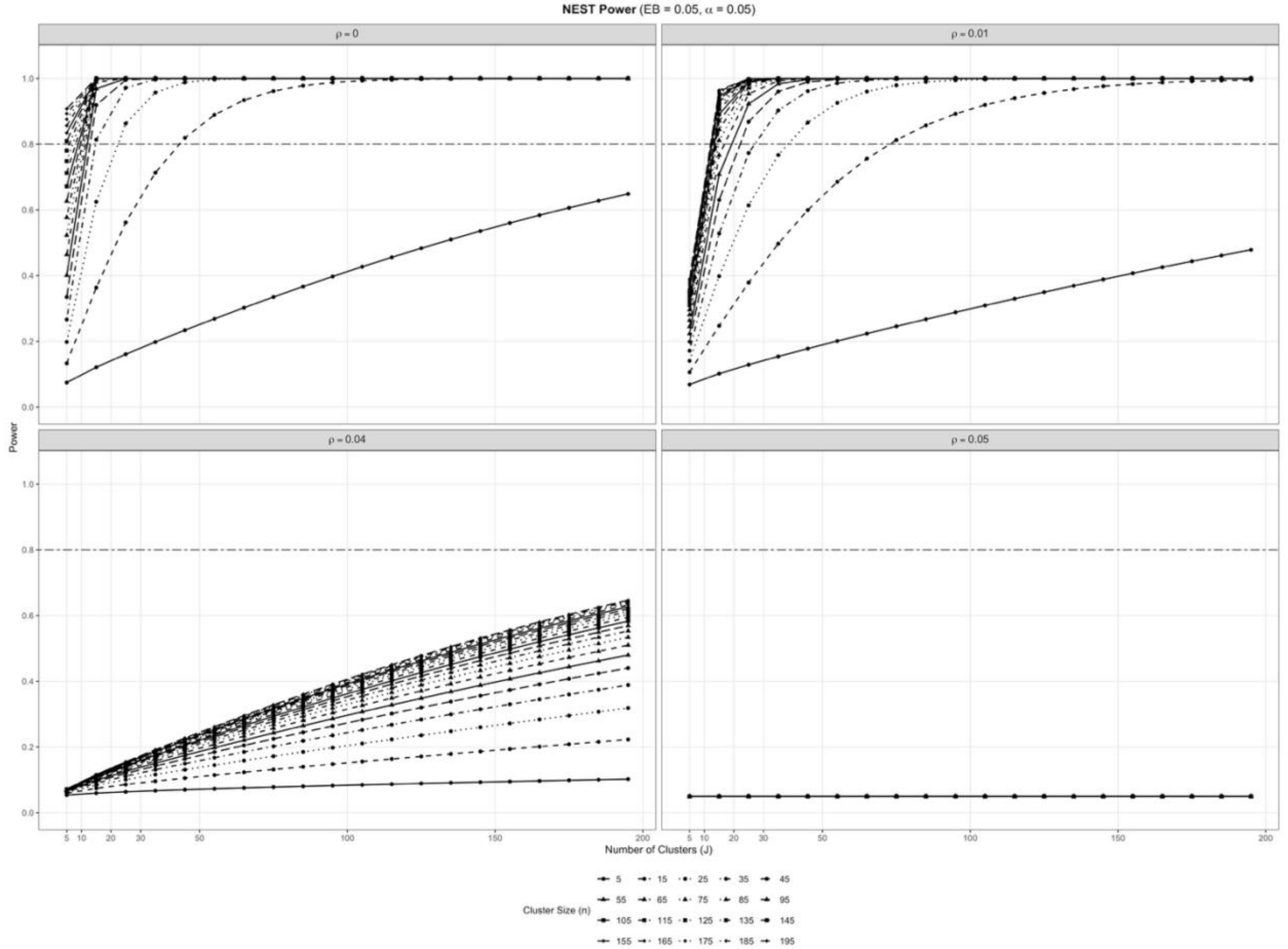

**Figure 2.**

*Power curves for NHST across number of clusters, by true ICC and cluster size*

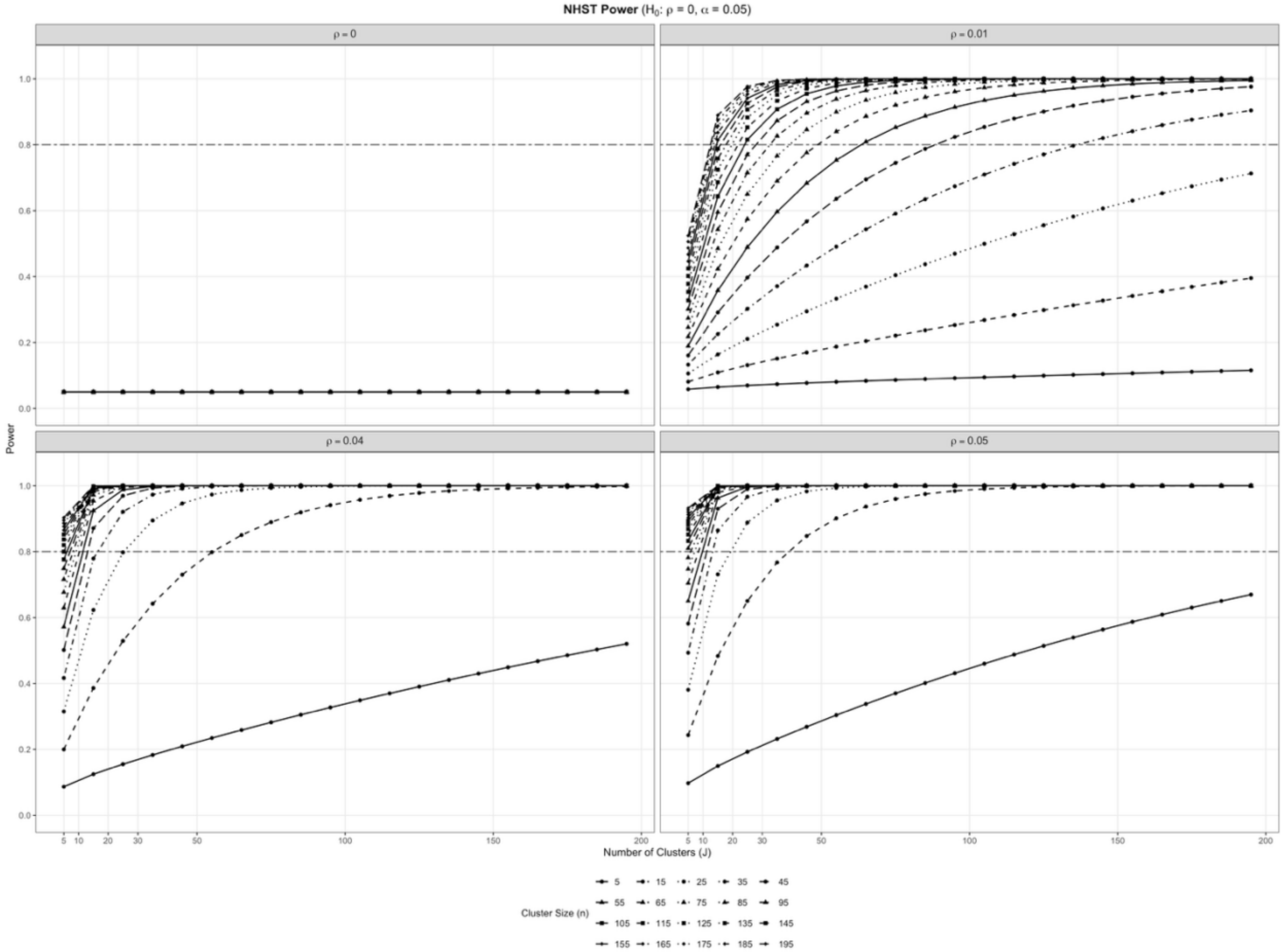

**References**

Altman, D. G., & Bland, J. M. (1995). Statistics notes: Absence of evidence is not evidence of absence. *British Medical Journal*, *311*(7003), 485. https://doi.org/10.1136/bmj.311.7003.485

Arnold, S. F. (1981). *The theory of linear models and multivariate analysis*. Wiley.

Beribisky, N., & Cribbie, R. A. (2026). A primer on equivalence (negligible effect) testing, *Psychological Methods*. Advance online publication. https://doi.org/10.1037/met0000800

Berkhof, J., & Kampen, J. K. (2004). Asymptotic effect of misspecification in the random part of the multilevel model. *Journal of Educational and Behavioral Statistics*, *29*(2), 201–218. https://doi.org/10.3102/10769986029002201

Blanca, M. J., Arnau, J., López-Montiel, D., Bono, R., & Bendayan, R. (2013). Skewness and kurtosis in real data samples. *Methodology, 9*(2), 78–84. https://doi.org/10.1027/1614-2241/a000057

Box, G. E. (1954). Some theorems on quadratic forms applied in the study of analysis of variance problems, I. Effect of inequality of variance in the one-way classification. *The annals of mathematical statistics, 25*(2), 290–302. https://doi.org/10.1214/aoms/1177728786

Burch, B. D. (2011a). Assessing the performance of normal-based and REML-based confidence intervals for the intraclass correlation coefficient. *Computational Statistics & Data Analysis*, *55*(2), 1018–1028. https://doi.org/10.1016/j.csda.2010.08.007

Burch, B. D. (2011b). Confidence intervals for variance components in unbalanced one-way random effects model using non-normal distributions. *Journal of Statistical Planning and Inference*, *141*(12), 3793–3807. https://doi.org/10.1016/j.jspi.2011.06.015

Casella, G., & Berger, R. L. (2002). *Statistical inference.* Duxbury Press.

Demetrashvili, N., Wit, E. C., & van den Heuvel, E. R. (2016). Confidence intervals for intraclass correlation coefficients in variance components models. *Statistical Methods in Medical Research*, *25*(5), 2359–2376. https://doi.org/10.1177/0962280214522787

Donner, A. (1986). A review of inference procedures for the intraclass correlation coefficient in the one-way random effects model. *International Statistical Review/Revue Internationale de Statistique, 54*(1), 67–82. https://doi.org/10.2307/1403259

Dorman, J. P. (2008). The effect of clustering on statistical tests: an illustration using classroom environment data. *Educational Psychology*, *28*(5), 583–595. https://doi.org/10.1080/01443410801954201

Dyer, N. G., Hanges, P. J., & Hall, R. J. (2005). Applying multilevel confirmatory factor analysis techniques to the study of leadership. *The Leadership Quarterly*, *16*(1), 149–167. https://doi.org/10.1016/j.leaqua.2004.09.009

Glass, G. V., Peckham, P. D., & Sanders, J. R. (1972). Consequences of failure to meet assumptions underlying the fixed effects analyses of variance and covariance. *Review of Educational Research*, *42*(3), 237–288. https://dx.doi.org/10.3102/00346543042003237

Goldstein, H., Browne, W., & Rasbash, J. (2002). Partitioning variation in multilevel models. *Understanding Statistics: Statistical Issues in Psychology, Education, and the Social Sciences*, *1*(4), 223–231. https://doi.org/10.1207/S15328031US0104_02

Hartung, J., & Knapp, G. (2005). On confidence intervals for the among-group variance in the one-way random effects model with unequal error variances. *Journal of Statistical Planning and Inference*, *127*(1-2), 157–177. https://doi.org/10.1016/j.jspi.2003.09.032

Harville, D. A., & Fenech, A. P. (1985). Confidence intervals for a variance ratio, or for heritability, in an unbalanced mixed linear model. *Biometrics*, *41*(1), 137–152. https://doi.org/10.2307/2530650

Hayes, A. F. (2006). A primer on multilevel modeling. *Human Communication Research, 32*, 385–410. https://dx.doi.org/10.1111/j.1468-2958.2006.00281.x

Huang, F. L. (2016). Alternatives to multilevel modeling for the analysis of clustered data. *Journal of Experimental Education, 84*(1), 175–196. https://doi.org/10.1080/00220973.2014.952397

Huang, F. L. (2018). Multilevel modeling myths. *School Psychology Quarterly, 33*(3), 492–499. https://doi.org/10.1037/spq0000272

Kasim, R. M., & Raudenbush, S. W. (1998). Application of Gibbs sampling to nested variance components models with heterogeneous within-group variance. *Journal of Educational and Behavioral Statistics*, *23*(2), 93–116. https://doi.org/10.3102/10769986023002093

Lakens, D. (2017). Equivalence test: A practical primer for t tests, correlations, and meta-analyses. *Social Psychological and Personality Science, 8*(4), 355–362. https://doi.org/10.1177/1948550617697177

Lakens, D., Scheel, A. M., & Isager, P. M. (2018). Equivalence testing for psychological research: A tutorial. *Advances in Methods and Practices in Psychological Science*, *1*(2), 259–269. https://doi.org/10.1177/2515245918770963

Leckie, G., French, R., Charlton, C., & Browne, W. (2014). Modeling heterogeneous variance–covariance components in two-level models. *Journal of Educational and Behavioral Statistics*, *39*(5), 307–332. https://doi.org/10.3102/1076998614546494

Lee, V. E. (2000). Using hierarchical linear modeling to study social contexts: The case of school effects. *Educational Psychologist*, *35*(2), 125–141. https://doi.org/10.1207/S15326985EP3502_6

Luke, D. A. (2004). *Multilevel modeling*, Sage. https://dx.doi.org/10.4135/9781412985147

McGraw, K. O., & Wong, S. P. (1996). Forming inferences about some intraclass correlation coefficients. *Psychological Methods, 1*(1), 30–46. https://doi.org/10.1037/1082-989X.1.1.30

McNeish, D. M., & Stapleton, L. M. (2016). The effect of small sample size on two-level model estimates: A review and illustration. *Educational Psychology Review*, *28*(2), 295–314. https://doi.org/10.1007/s10648-014-9287-x

McNeish, D., Stapleton, L. M., & Silverman, R. D. (2017). On the unnecessary ubiquity of hierarchical linear modeling. *Psychological Methods*, *22*(1), 114–140. https://doi.org/10.1037/met0000078

Meyners, M. (2012). Equivalence tests – A review. *Food Quality and Preference*, *26*(2), 231–245. https://doi.org/10.1016/j.foodqual.2012.05.003

Peugh, J. L. (2010). A practical guide to multilevel modeling. *Journal of School Psychology, 48*(1), 85–112. https://doi.org/10.1016/j.jsp.2009.09.002

Pituch, K. A., Murphy, D. L., & Tate, R. L. (2010). Three-level models for indirect effects in school- and class-randomized experiments in education. *The Journal of Experimental Education, 78*(1), 60–95. https://doi.org/10.1080/00220970903224685

Raudenbush, S., & Bryk, A. (2002). *Hierarchical linear models: Applications and data analysis methods* (2nd ed.). Sage.

Raudenbush, S. W., Bryk, A., Cheong, Y. F., Congdon, R., & du Toit, M. (2004). *HLM6: Hierarchical linear and nonlinear modeling*. Scientific Software International.

Raykov, T. (2011). Intraclass correlation coefficients in hierarchical designs: Evaluation using latent variable modeling. *Structural Equation Modeling: A Multidisciplinary Journal, 18*(1), 73–90. https://doi.org/10.1080/10705511.2011.534319

Rogers, J. L., Howard, K. I., & Vessey, J. T. (1993). Using significance tests to evaluate equivalence between two experimental groups. *Psychological Bulletin, 113*(3), 553–565. https://doi.org/10.1037/0033-2909.113.3.553

Rusticus, S. A. & Lovato, C. Y. (2014). Impact of sample size and variability on the power and Type I error rates of equivalence tests: A simulation study. *Practical Assessment, Research, and Evaluation, 19*(1), 11. http://doi.org/10.7275/4s9m-4e81

Schuirmann, D. J. (1987). A comparison of the two one-sided tests procedure and the power approach for assessing the equivalence of average bioavailability. *Journal of Pharmacokinetics and Biopharmaceutics, 15*(6), 657–680. https://doi.org/10.1007/BF01068419

Seaman, M. A., & Serlin, R. C. (1998). Equivalence confidence intervals for two-group comparisons of means. *Psychological Methods, 3*(4), 403–411. https://doi.org/10.1037/1082-989X.3.4.403

Shrout, P. E., & Fleiss, J. L. (1979). Intraclass correlations: uses in assessing rater reliability. *Psychological Bulletin*, *86*(2), 420–428. https://doi.org/10.1037/0033-2909.86.2.420

Snijders, T. A. B., & Bosker, R. J. (2012). *Multilevel analysis: An introduction to basic and advanced multilevel modeling* (2nd ed.). Sage.

Swiger, L. A., Harvey, W. R., Everson, D. O., & Gregory, K. E. (1964). The variance of intraclass correlation involving groups with one observation. *Biometrics*, *20*(8). 818–826, https://doi.org/10.2307/2528131

Thomas, J. D., & Hultquist, R. A. (1978). Interval estimation for the unbalanced case of the one-way random effects model. *The Annals of Statistics, 6*(3), 582–587. https://doi.org/aos/1176344202

Thomas, S. L., & Heck, R. H. (2001). Analysis of large-scale secondary data in higher education research: Potential perils associated with complex sampling designs, *Research in Higher Education, 42*(5), 517–540. https://dx.doi.org/10.1023/A:1011098109834

Tryon, W. W. (2001). Evaluating statistical difference, equivalence, and indeterminacy using inferential confidence intervals: An integrated alternative method of conducting null hypothesis statistical tests. *Psychological Methods, 6*(4), 371–386. https://doi.org/10.1037/1082-989X.6.4.371

Ukoumunne, O. C., Davison, A. C., Gulliford, M. C., & Chinn, S. (2003). Non-parametric bootstrap confidence intervals for the intraclass correlation coefficient. *Statistics in Medicine*, *22*(24), 3805–3821. https://doi.org/10.1002/sim.1643

Verbeke, G., & Lesaffre, E. (1997). The effect of misspecifying the random-effects distribution in linear mixed models for longitudinal data. *Computational Statistics & Data Analysis*, *23*(4), 541–556. https://doi.org/10.1016/S0167-9473(96)00047-3

Wald, A. (1940). A note on the analysis of variance with unequal class frequencies. *The Annals of Mathematical Statistics*, *11*(1), 96–100. https://doi.org/10.1214/aoms/1177731946

Wellek, S. (2010). *Testing statistical hypotheses of equivalence and noninferiority* (2nd ed.). Chapman & Hall/CRC.

Westlake, W. J. (1976). Symmetrical confidence intervals for bioequivalence trials. *Biometrics, 32*(4), 741–744. https://doi.org/10.2307/2529259

Westlake, W. J. (1981). Bioequivalence testing--a need to rethink. *Biometrics*, *37*(3), 589–594. https://doi.org/10.2307/2530573

Wilkinson, L., & The Task Force on Statistical Inference. (1999). Statistical methods in psychology journals: Guidelines and explanations. *American Psychologist, 54*(8), 594–604. https://doi.org/10.1037/0003-066X.54.8.594

**Appendix A**

**Power for NEST for the ICC**

For the balanced design, the Eq. (3.10) can be written as:

$$\frac{F_{\text{obs}}}{\left(1+n\frac{\rho}{(1-\rho)}\right)} \sim F_{J-1,N-J}.$$

In our NEST framework, $H_0$ is rejected if the upper bound of the ICC is less than EB. Therefore, from the Eq. (4.4),

$$\frac{F_{\text{obs}} - F_{\alpha,J-1,N-J}}{F_{\text{obs}} + (n-1)F_{\alpha,J-1,N-J}} < \text{EB}.$$

This can be rewritten as

$$F_{\text{obs}} - F_{\alpha,J-1,N-J} < \text{EB}\left(F_{\text{obs}} + (n-1)F_{\alpha,J-1,N-J}\right)$$

$$(\text{EB})F_{\text{obs}} < F_{\alpha,J-1,N-J} + \text{EB}(n-1)F_{\alpha,J-1,N-J}$$

$$F_{\text{obs}} < F_{\alpha,J-1,N-J}\left(\frac{1+\text{EB}(n-1)}{1-\text{EB}}\right). \tag{A.1}$$

Eq. (A.1) is the condition of rejecting $H_0$. Given that the Eq. (3.9) is:

$$F_{\text{obs}} \sim \left(1+n\frac{\rho}{(1-\rho)}\right)F_{J-1,N-J}, \tag{A.2}$$

by substituting Eq. (A.2) to Eq. (A.1),

$$\frac{(n-1)\rho+1}{1-\rho}F_{J-1,N-J} < F_{\alpha,J-1,N-J} \cdot \left(\frac{1+\text{EB}(n-1)}{1-\text{EB}}\right)$$

$$F_{J-1,N-J} < \frac{1-\rho}{(n-1)\rho+1} \cdot F_{\alpha,J-1,N-J} \cdot \frac{1+\text{EB}(n-1)}{1-\text{EB}}. \tag{A.3}$$

Therefore, power ($\pi$) can be defined as the probability of Eq. (A.3):

$$\pi = \Pr\left(F_{J-1,N-J} < \frac{1-\rho}{(n-1)\rho + 1} \cdot F_{\alpha,J-1,N-J} \cdot \frac{1 + \mathrm{EB}(n-1)}{1 - \mathrm{EB}}\right).$$

**Appendix B**

**Extension to examine conditional ICCs in random-intercept models with covariates**

When the covariates are included, Eqs. (3.1) and (3.2) can be written follows:

$$\text{Level 1: } Y_{ij} = \beta_{0j} + \gamma_W\left(X_{ij} - \bar{X}_{.j}\right) + r_{ij,c}, \quad r_{ij,c} \sim N(0, \sigma_c^2)$$

$$\text{Level 2: } \beta_{0j} = \gamma_{00} + \gamma_B B_j + u_{0j,c}, \quad u_{0j,c} \sim N(0, \tau_{00,c})$$

where $B_j$ denotes level 2 covariates, and $X_{ij} - \bar{X}_{.j}$ is the group-mean centered level 1 covariate. The conditional ICC (cICC, $\rho_c$) is as follows (Raudenbush & Bryk, 2002)

$$\rho_c = \frac{\tau_{00,c}}{\tau_{00,c} + \sigma_c^2}. \quad \text{(B.1)}$$

Let $R_{\mathrm{B}}^2$ denote the proportion of between-group variance explained by level 2 covariates, and let $R_{\mathrm{W}}^2$ denote the proportion of within-group variance explained by level 1 covariates. Then,

$$\tau_{00,c} = \tau_{00}(1 - R_{\mathrm{B}}^2) \quad \text{(B.2)}$$

$$\sigma_c^2 = \sigma^2(1 - R_{\mathrm{W}}^2) \quad \text{(B.3)}$$

and therefore,

$$\rho_c = \frac{\tau_{00}(1 - R_{\mathrm{B}}^2)}{\tau_{00}(1 - R_{\mathrm{B}}^2) + \sigma^2(1 - R_{\mathrm{W}}^2)} = \frac{\rho(1 - R_{\mathrm{B}}^2)}{\rho(1 - R_{\mathrm{B}}^2) + (1 - \rho)(1 - R_{\mathrm{W}}^2)}. \quad \text{(B.4)}$$

Eq. (B.4) shows that the unconditional and conditional ICCs are generally different.

*Case 1) When the covariates explain only between-group variation*

If the model only includes level 2 covariates, the covariates explain only between-group variation (i.e., $R_{\mathrm{W}}^2 = 0$). In this case, following Eq. (B.4),

$$\rho_c = \frac{\tau_{00}(1 - R_{\mathrm{B}}^2)}{\tau_{00}(1 - R_{\mathrm{B}}^2) + \sigma^2} = \frac{\rho(1 - R_{\mathrm{B}}^2)}{\rho(1 - R_{\mathrm{B}}^2) + (1 - \rho)} = \frac{\rho(1 - R_{\mathrm{B}}^2)}{1 - \rho R_{\mathrm{B}}^2} \le \rho. \quad \text{(B.5)}$$

Since $\rho \leq 1$, it implies $\rho R_{\mathrm{B}}^2 \leq R_{\mathrm{B}}^2$, and hence $1 - \rho R_{\mathrm{B}}^2 \geq 1 - R_{\mathrm{B}}^2$. Given that the ratio of $\frac{1-R_{\mathrm{B}}^2}{1-\rho R_{\mathrm{B}}^2}$ is at most 1, $\rho_c \leq \rho$ can be derived. Therefore, in this case, if the unconditional ICC ($\rho$) is negligible, it also implies that the conditional ICC ($\rho_c$) is negligible under the same EB.

*Case 2) When the covariate explains only within-group variation*

When level 1 covariates explain within-cluster variation (i.e., $R_{\mathrm{B}}^2 = 0$ and $R_{\mathrm{W}}^2 > 0$), then the conditional ICC can be larger than unconditional ICC.

$$\rho_c = \frac{\tau_{00}}{\tau_{00} + \sigma^2(1 - R_{\mathrm{W}}^2)} > \rho \qquad (\mathrm{B}.6)$$

Thus, in this case, the negligibility of the unconditional ICC does not ensure that of the conditional ICC. If the goal is to guarantee that the conditional ICC is below a prespecified EB, the unconditional ICC must therefore be tested against a stricter EB.

Let $R_{\mathrm{W,max}}^2$ denote the largest proportion of within-cluster variance that the level 1 covariates are expected to explain (i.e., $R_{\mathrm{W}}^2 < R_{\mathrm{W,max}}^2$). Then we need to set the EB ($\delta$) to be an *adjusted* form ($\delta_{\mathrm{adj}}$) to ensure $\rho_c \leq \delta$ if $\rho \leq \delta_{\mathrm{adj}}$. First, for fixed $\rho$, $\rho_c$ in Eq. (B.6) increases as $R_{\mathrm{W}}^2$ increases. Thus, for any given $\rho$, $\rho_c$ is maximized when $R_{\mathrm{W}}^2 = R_{\mathrm{W,max}}^2$. Second, assuming $R_{\mathrm{W}}^2 = R_{\mathrm{W,max}}^2$, $\rho_c$ is the monotonically increasing function:

$$\rho_c = f(\rho) = \frac{\rho}{\rho + (1-\rho)(1 - R_{\mathrm{W,max}}^2)}$$

$$f'(\rho) = \frac{(1 - R_{\mathrm{W,max}}^2)}{[\rho + (1-\rho)(1 - R_{\mathrm{W,max}}^2)]^2} > 0\ .$$

Taken together, over all pairs ($\rho$, $R_{\mathrm{W}}^2$) with $\rho \leq \delta_{\mathrm{adj}}$ and $R_{\mathrm{W}}^2 = R_{\mathrm{W,max}}^2$ is the worst case. Because $\delta_{\mathrm{adj}}$ was chosen so that $f(\delta_{\mathrm{adj}}) = \delta$, $\rho_c$ cannot exceed $\delta$ for any model in this range, and it equals $\delta$ only in the worst case:

$$f(\delta_{\text{adj}}) = \frac{\delta_{\text{adj}}}{[\delta_{\text{adj}} + (1 - \delta_{\text{adj}})(1 - R^2_{\text{W,max}})} = \delta\,. \tag{B.7}$$

Rearranging Eq. (B.7) yields

$$\delta_{\text{adj}} = \frac{\delta(1 - R^2_{\text{W,max}})}{(1 - \delta R^2_{\text{W,max}})}. \tag{B.8}$$

Therefore, whenever the ρ falls below $\delta_{\text{adj}}$, $\rho_c$ remains below δ for all models whose level 1 covariates explain at most $R^2_{\text{W,max}}$ of the within-cluster variance. Prior empirical studies can guide the choice of a value for $R^2_{\text{W,max}}$, which represents how much within-cluster variance may be explained by covariates.

*Case 3) When the covariates explain variation at both levels*

Combining Cases 1 and 2, the conditional variance components are $\tau_{00,c} = \tau_{00}(1 - R^2_{\text{B}})$ and $\sigma^2_c = \sigma^2(1 - R^2_{\text{W}})$. As seen previously, the two mechanisms pull $\rho_c$ in opposite directions. While explained between-group variance lowers $\rho_c$, explained within-group variance raises it, and the two effects offset exactly when $R^2_{\text{B}} = R^2_{\text{W}}$. Moving from ρ to $\rho_c$ multiplies the ratio $\tau_{00}/\sigma^2$ by $(1 - R^2_{\text{B}})/(1 - R^2_{\text{W}})$, so the $\rho_c$ exceeds the ρ when $R^2_{\text{W}} > R^2_{\text{B}}$. Fortunately, the separate adjustment is not needed for Case 3. For a fixed $R^2_{\text{W}}$, the conditional ICC decreases as $R^2_{\text{B}}$ increases; thus, the largest possible $\rho_c$ occurs when $R^2_{\text{B}} = 0$, which is actually the same with Case 2. Therefore, the adjusted EB ($\delta_{\text{adj}}$) derived for Case 2 also covers this case.

## Appendix C

### R code for the illustration

```
library(EdSurvey)
library(dplyr)

elcs <- readECLS_K2011(
  path = getwd(),
  filename = “childK5p.dat”,  # this file should be inside the
directory
  layoutFilename = “ECLSK2011_K5PUF.sps”,  # this file should
be inside the directory
  forceReread = FALSE,
  verbose = TRUE
)

id_vars <- c(“childid”, “s7_id”)

attitude <- c(“c7lkread”)

all_vars_to_extract_lower <- c(id_vars, attitude)

my_data <- EdSurvey::getData(data = elcs, varnames =
all_vars_to_extract_lower)



## 1. Descriptive statistics (N, J)
num_individuals <- nrow(my_data)  # N
num_groups <- length(unique(my_data$s7_id))  # J
group_sizes <- table(my_data$s7_id)  # “s7_id” is the cluster-
level ID
avg_group_size <- (1/(num_groups - 1)) * (num_individuals -
(sum(table(my_data$s7_id)^2)) / num_individuals)  # average
cluster size corrected for variance (n_0)

print(paste(“N (number of individuals):”, num_individuals))
print(paste(“k (number of clusters):”, num_groups))
print(paste(“n_0 (average cluster size corrected for
variance):”, avg_group_size))



## 2. Compute the ICC
library(lme4)
my_data$c7lkread_num <- as.numeric(my_data$c7lkread)
```

```
null_model <- lmer(c7lkread_num ~ 1 + (1 | s7_id), data =
my_data)

summary(null_model)

variances <- as.data.frame(VarCorr(null_model))

school_variance <- variances[1, "vcov"]

residual_variance <- variances[2, "vcov"]

icc <- school_variance / (school_variance + residual_variance)

print(paste("ICC:", round(icc, 3)))



## 3. Compute MS_W, MS_B, and F statistic
anova_data <- my_data[, c("s7_id", "c7lkread_num")]  #"s7_id"
is a level-2 ID, "c7lkread_num" is an outcome variable
anova_data <- na.omit(anova_data)

anova_data$s7_id <- as.factor(anova_data$s7_id)

aov_model <- aov(c7lkread_num ~ s7_id, data = anova_data)

aov_summary <- summary(aov_model)

ms_between <- aov_summary[[1]]["Mean Sq"][1, ]
ms_within <- aov_summary[[1]]["Mean Sq"][2, ]

print(paste("MS_between: ", round(ms_between, 4)))
print(paste("MS_within: ", round(ms_within, 4)))

f_statistic <- ms_between / ms_within

print(paste("F statistic: ", round(f_statistic, 4)))



## 4. Compute EB using DEFF (optional)
DEFF <- 1.1
EB <- (DEFF-1)/(avg_group_size-1)

print(paste("EB: ", round(EB, 4)))
```

```
## 5. Critical value for lower 100(a)% tail of the F-
distribution
# df1: df_between (numerator)
# df2: df_within (denominator)
df1 <- num_groups – 1
df2 <- num_individuals – num_groups
alpha <- 0.05
f_lower <- qf(alpha, df1 = df1, df2 = df2)

print(paste("F_L (Critical value at", alpha, "lower tail):",
round(f_lower, 4)))



## 6. Upper bound of CI
rho_upper <- (f_statistic - f_lower)/(f_statistic +
(avg_group_size - 1)*f_lower)
print(paste("Upper bound of CI:", round(rho_upper, 4)))
```